\RequirePackage{fix-cm}
\documentclass[smallextended]{svjour3}
\smartqed

\usepackage{graphicx}
\usepackage[utf8]{inputenc}
\usepackage{graphicx}
\usepackage{float}
\usepackage{amsmath}
\usepackage[utf8]{inputenc}
\usepackage[english]{babel}
\DeclareUnicodeCharacter{2212}{-}
\DeclareUnicodeCharacter{2191}{\uparrow}
\DeclareUnicodeCharacter{200E}{}
\usepackage{array, makecell}
\usepackage{boldline}
\usepackage{colortbl}
\begin{document}

\title{Tunable Multistate Terahertz Switch Based on Multilayered Graphene Metamaterial}

\author{Dip Sarker$^{1,2,\dagger}$ \and
	Partha Pratim Nakti$^{1,\dagger}$ \and
	Md Ishfak Tahmid$^{1,2}$ \and Md Asaduz Zaman Mamun$^{1}$ \and Ahmed Zubair$^{2}$
}

\institute{\at
	$^1$Department of Electrical and Electronic Engineering,  Shahjalal University of Science and Technology, Sylhet, Bangladesh. 
	\at 
	$^2$Department of Electrical and Electronic Engineering, Bangladesh University of Engineering and Technology, Dhaka, Bangladesh. 
	\at 
	$\dagger$These authors contributed equally to this work. 
	\and
    A.\,Zubair \at
	\email{ahmedzubair@eee.buet.ac.bd}
	}
\date{Received: date / Accepted: date}
\maketitle
\begin{abstract}
We proposed plasmonic effect based narrow band tunable terahertz switches consisting of multilayered graphene metamaterial. Though several terahertz optical switches based on metamaterials were previously reported, these switches had complicated fabrication processes, limited tunability, and low modulation depths. We designed and simulated ingenious four and eight state terahertz optical switch designs that can be functional for multimode communication or imaging using the finite-difference time-domain simulation technique. The plasmonic bright modes and transparency regions of these structures were adjusted by varying the chemical potential of patterned graphene layers via applying voltage in different layers. The structures exhibited high modulation depth and modulation degree of frequency, low insertion loss, high spectral contrast ratio, narrow bandwidth, and high polarization sensitivity. Moreover, our proposed simple fabrication process will make these structures more feasible compared to previously reported terahertz switches. The calculated modulation depths were 98.81\% and 98.71\%, and maximum modulation degree of frequencies were $\sim$61\% and $\sim$29.1\% for four and eight state terahertz switches, respectively. The maximum transmittance in transparency regions between bright modes and the spectral contrast ratio were enumerated to be 95.9\% and $\sim$96\%, respectively. The maximum insertion losses were quite low with values of 0.22 dB and 0.33 dB for four and eight state terahertz switches, respectively. Our findings will be beneficial in the development of ultra-thin graphene-based multistate photonic devices for digital switching, sensing in terahertz regime.

\end{abstract}

\section{Introduction}
Due to its distinctive properties, EM waves in the terahertz (THz) frequency regime have inspired tremendous academic and technological interest because of their various applications including wireless communication, chemical detection, and biomedical imaging~\cite{Tonouchi2007,Nagatsuma2016,Zhang2021}. Over the last decade, many techniques, including those based on active metamaterials and two-dimensional materials have been studied to control electromagnetic wave propagation from radio to THz frequency~\cite{Zaytsev2019,Tao2011,Wang2018}. Graphene, a highly conductive single-atom-thick material with hexagonal shape of carbon atom, exhibits long-lived plasmon excitation as well as outstanding low loss, and customizable properties. These characteristics make graphene metamaterial based optical switches useful in switching scheme with enhanced efficiency. Consequently, several experimental and theoretical works on graphene metamaterial based infrared optical switches, where plasmonics effect can eliminate the opaque effect of the medium to the electromagnetic wave using surface plasmon polariton (SPP), were reported~\cite{Sarker2021,Zhang2020OE,zhang2008plasmon,hu2016tunable,Liu2020,granpayeh2018tunable,li2020ultracompact,luo2016flexible}. In particular, optical switches are essential elements in various applications such as contemporary communication system~\cite{Reichel2018} and signal measurement~\cite{Song2017}.

A transmission transparency of 90\% and 70\% between two bright modes was reported by Tang \textit{et al.}~\cite{Tang2019} and Ou \textit{et al.}~\cite{Ou2020}, respectively for THz optical switch using metal-based metamaterial on dielectric and silicon substrate. However, tuning of resonant frequencies in these structures was not possible without modifying the physical structures. In contrast, the surface conductivity of the graphene metamaterials can be modulated directly without changing structural parameters by applying voltage. Habib \textit{et al.}~\cite{Habib2018} introduced gold strips on top of graphene and the structure had a spectral contrast ratio of 82\% at mid infrared region. Li \textit{et al.}~\cite{li2018graphene} reported a structure with left-right electrodes instead of conventional top-bottom electrodes. Though such designs allowed the transmittance of terahertz (THz) signal to be regulated by a voltage source, the performance parameter modulation depth (MD) was low (89\%) compared to previous studies. MD of 92\% was achieved by Khazaee \textit{et al.}~\cite{granpayeh2018tunable} at mid-infrared frequency regime but transmission intensity between two bright modes was very poor ($\sim$70\%) for an optical switch. Cascaded two state switch elements allowed for more connection through concatenating additional switching states; however, achieving high extinction ratios in such two state switch elements had proven problematic than multistate switch~\cite{Stabile2012}. Although there were a very limited reports of four state optical switch in the literature, eight state switch in THz regime is yet to be explored. Moreover, there is scope for significant improvement in switch performance for THz communication and sensing applications in terms of performance parameters.

In this paper, we proposed high performance four and eight state narrow band THz optical switches based on two and three patterned layer of graphene sheets, respectively, utilizing the concept of SPP phenomenon. The optical switches were realized by tuning chemical potential in the patterned graphene layers on polymethyl methacrylate (PMMA) and consequently producing multiple plasmonic modes. We optimized the designs and evaluated the performance of these devices using finite-difference time-domain (FDTD) method. Tuning of the chemical potentials was accomplished by adjusting surface conductivity of graphene via voltage sources. Additionally, we performed polarization sensitivity analysis of our proposed narrow band THz switch and proposed a viable fabrication process. A comparative analysis of the performance parameters of our switches and recently demonstrated metamaterial based optical switches were conducted. This study will pave a new way to design and fabricate optical switches.

\section{Methodology}
\label{sec:SYSTEM}
\begin{figure}[H]
\centering\includegraphics[width=\textwidth]{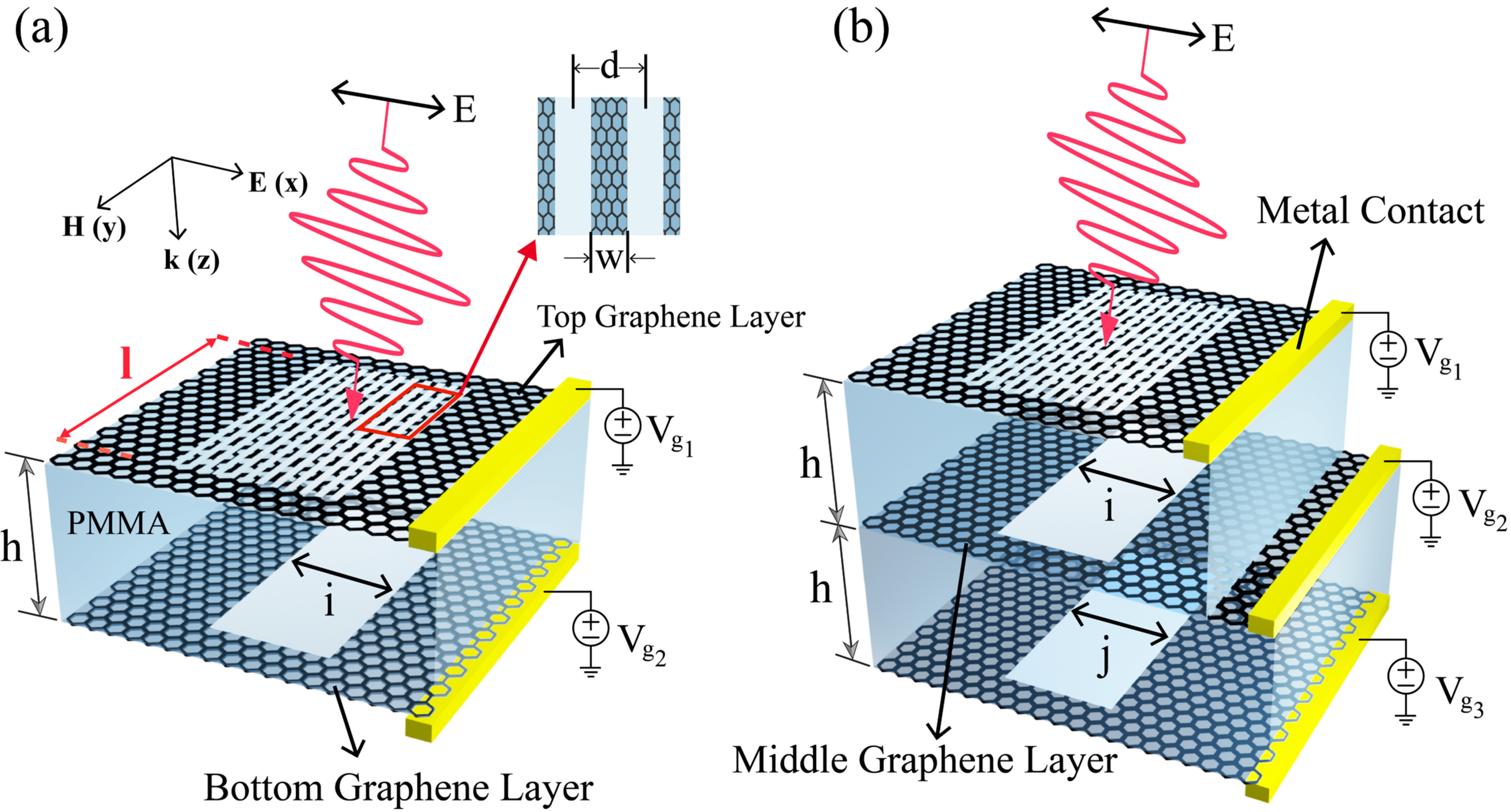}
\caption{(a) Illustration of proposed four state graphene metamaterial based THz switch under x-polarized THz electric field propagating along z direction. Here, d = 20 nm, w = 10 nm, l = 2.5 $\mu$m, length and width of PMMA = 3 $\mu$m, h = 2 $\mu$m, and i = 150 nm. (b) Three dimension view of eight state graphene metamaterial based THz switch with geometrical parameters: h = 2 $\mu$m, l = 2.5 $\mu$m, length and width of PMMA = 3 $\mu$m, i = j = 150 nm, d = 20 nm, and w = 10 nm. ${\mbox{V}}_{\mbox{{g}}_{\mbox{n}}}$ is the gate voltage used to tune the chemical potential of n-th graphene layer.}
\label{fig:1}
\end{figure}

Patterned graphene monolayer on both sides of PMMA substrate constituted our four state THz switch design as can be seen in Fig.~\ref{fig:1}(a). The monolayer graphene can be grown using chemical vapor deposition (CVD) process. The helium ion beam lithography (HIBL) technique can be introduced to create patterns on the CVD grown graphene layers. The top graphene layer consists of a periodic array of strips of graphene nanoribbon and air hole. The bottom layer consists of a single rectangular air hole which can be created using HIBL. A suggested fabrication process is presented in \textcolor{blue}{Supplementary Information}. The length and width of  PMMA substrate were set to be 3 $\mu$m. The thickness of PMMA layers was chosen 2 $\mu$m to avoid near field effect (See \textcolor{blue}{Fig. S3} in \textcolor{blue}{Supplementary Information} for details) and Fabry-Perot multi-reflections~\cite{Sarker2021}. The nanoribbon strip width, w of the top graphene layer was 10 nm, with pitch distance, d of 20 nm, and air hole length, l of 2.5 $\mu$m. Detailed analyses of resonant frequency dependency on w and d were performed elsewhere~\cite{Sarker2021}. The rectangular air hole of the bottom graphene layer width, i was set to be 150 nm. The graphene layers were connected to voltage sources via metal contacts to vary the chemical potential, $\mu_{c_n}$ of the corresponding $n$-th graphene layers.

Three layers of patterned graphene on PMMA substrate comprised our proposed eight state THz switch design as illustrated in Fig.~\ref{fig:1}(b). Similar fabrication technique can be used to form the patterns on three graphene layers. We provided a detailed fabrication method in \textcolor{blue}{Supplementary Information}. Two PMMA substrates had identical height (h) of 2 $\mu$m. The patterned top graphene layer had similar parameters as four state switch's top graphene layer. The middle and bottom graphene layers had a rectangular air hole having width of i = j = 150 nm. Voltage sources were connected through metal contacts to allow the tuning of $\mu_{c_n}$.

The study was conducted using 3D FDTD numerical analysis technique. Periodic boundary conditions were employed in the x- and y-directions. In the positive and negative z-directions, the steep angle perfectly matched layer (PML) boundary condition was adopted. To avoid light reflection from the PML border, the thickness of the PML layer was designed to be larger than the source's peak wavelength. For our simulation, we used a non-uniform conformal mesh. A plane wave THz light was incident from the top along the z-direction. The transmission spectra were recorded by a power monitor at the bottom of the structure. For all simulations, the temperature was set to be 300 K and the relaxation time was set to be 2 ps.

The well-known Kubo's formulation was used to simulate the surface conductivity of graphene which consists of intraband and interband transitions of electron~\cite{hanson2008dyadic,casiraghi2007rayleigh,Li2018}. The detailed formulation can be found elsewhere~\cite{Sarker2021}.  The conductivity of graphene, $\sigma$ can be obtained by~\cite{Rouhi2012},
\begin{equation}
    \sigma = \frac{ie^{2}E_f}{\pi \hbar^2(\omega+i\tau^{-1})}.
    \label{eq1}
\end{equation}
where $\tau = \mu E_f / e{V_{F}}^{2}$ is the carrier relaxation time. $\mu, \omega, e, \hbar, E_{f}$, and $V_{F}$  are the mobility of graphene, the angular frequency of the incident
light, the electron charge, the reduced Planck constant, the Fermi level of graphene, and the Fermi velocity, respectively.
\section{Results and Discussion}
\begin{figure}[H]
\centering\includegraphics[width=\textwidth]{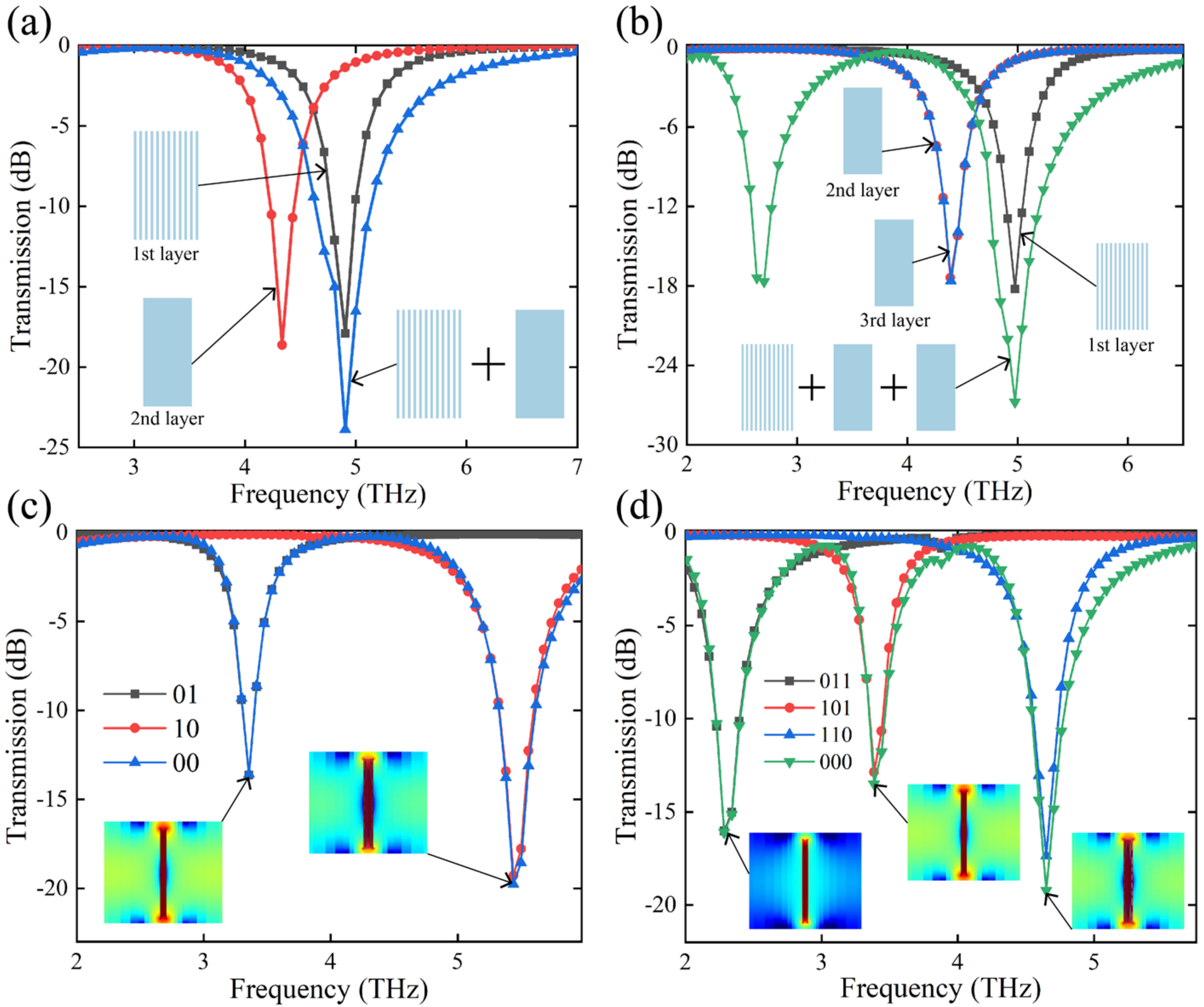}
\caption{(a) Transmission spectra of GNR nanoribons, rectangular air hole in graphene, and two patterned graphene metamaterial. Here, the chemical potential of graphene was set to 0.4 eV. (b) Transmission spectra of GNR strips, rectangular air hole in graphene, and three patterned graphene metamaterial. Transmission spectra of (c) four and (d) eight state switching scheme using different chemical potentials. Dominated dark plasmonic modes in graphene layer are shown in inset.}
\label{fig2}
\end{figure}

Interaction between incoming plane wave and patterned graphene layer resulted in plasmon resonance on the surface of graphene layer which attenuates light by scattering. Moreover, transmission of light through the structure was blocked at a certain resonant frequency because of strong plasmonic interaction. Hence, each patterned graphene layer on PMMA act as an attenuating medium resulting in a plasmonic bright mode at a certain resonant frequency. The transmission spectra formed in different GNR metamaterial layers were depicted in Fig.~\ref{fig2}. The GNR nanoribbon gave rise to the plasmonic bright mode (black Lorentz curve) as shown in Fig.~\ref{fig2}(a). The GNR layer with rectangular air hole produced a plasmonic bright mode (red Lorentz curve). When the two structures were merged in a structure as top and bottom layers, coupling of the bright modes blue-shifted the plasmonic bright mode originated from the air hole as can be seen in Fig.~\ref{fig2}(a) (blue curve). Here, the chemical potential of graphene layer was set to be 0.4 eV for all layers. Similarly, three engineered layers of graphene illustrated plasmonic effects as shown in Fig.~\ref{fig2}(b). To separate and produce such two bright plasmonic modes from coupled Lorentz curve, patterned graphene layers at different chemical potentials by varying surface conductivity can be utilized as shown in Fig.~\ref{fig2}(c) and (d).
The design of two and three patterned graphene layers were utilized to create two and three stymies of light transmission and formed narrow transparency windows between them (See \textcolor{blue}{Fig. S2} of \textcolor{blue}{Supplementary Information}). It is apparent from the reflectance and absorption spectra for four and eight state THz switches that reflection and absorption in these transparency windows were very low (\textcolor{blue}{Fig. S2} of \textcolor{blue}{Supplementary Information}).
\subsection{Four State Switching Scheme}

\begin{figure}[H]
\centering\includegraphics[width=\textwidth]{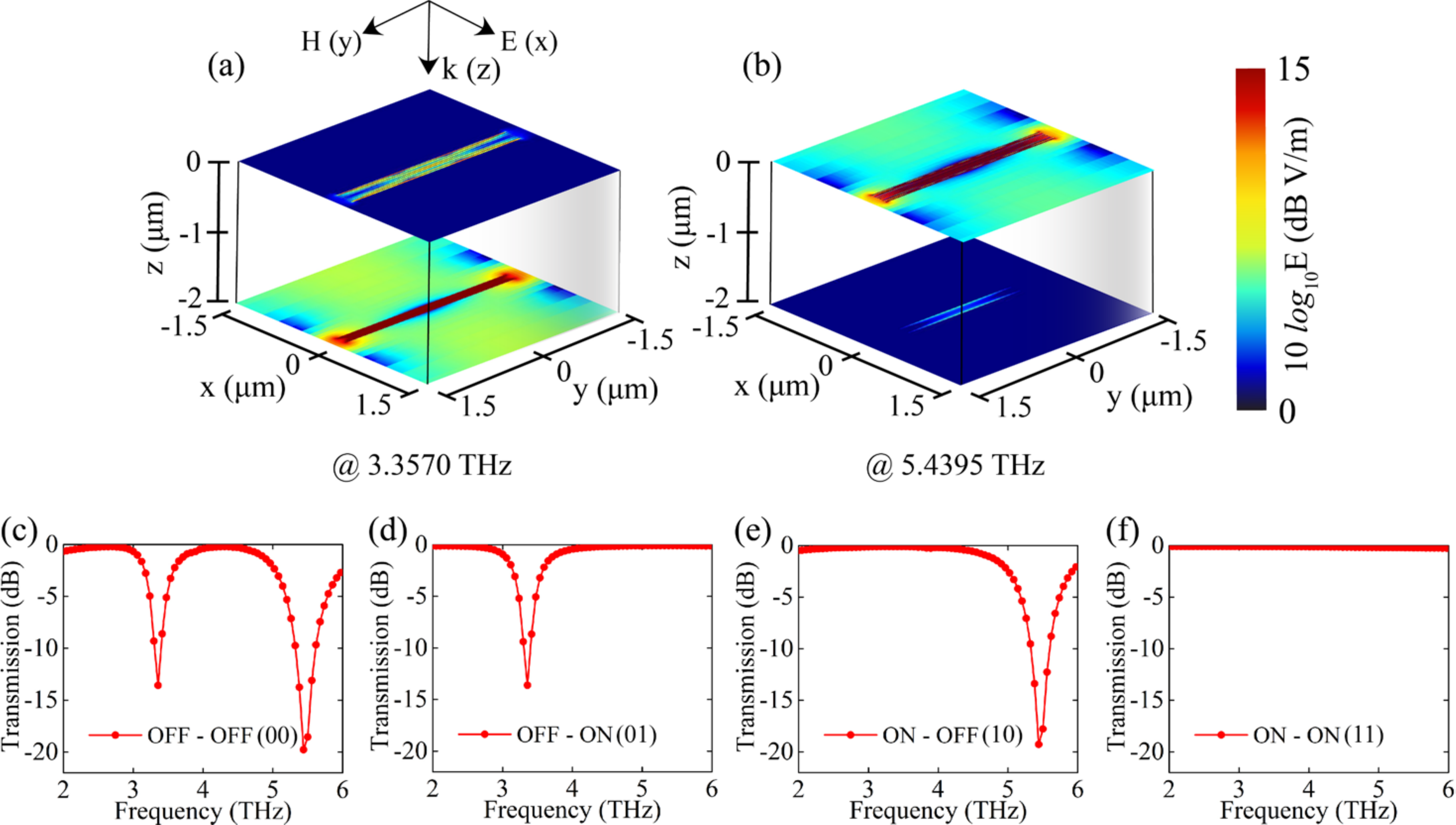}
\caption{The spatial E-field distribution of top and bottom patterned graphene layers at resonant frequencies of (a) 3.357 THz, and (b) 5.4395 THz. Transmission spectra of four state switch at different states: (c) “00” state, (d) “01” state, (e) “10” state, and (f) “11” state. The E-field distributions are shown in log scale.}
\label{fig3}
\end{figure}

To achieve four state switching scheme, we utilized localized plasmon resonance phenomenon in two stacked patterned graphene layers. In order to get a better comprehension of plasmonic effect, the field distributions of two dips are presented in Figs.~\ref{fig3}(a) and (b). The light field was localized primarily around the bottom patterned graphene layer at resonant frequency of 3.357 THz, as shown in Fig.~\ref{fig3}(a). Similarly, the top patterned graphene layer had localized electric field at 5.4395 THz creating the bright plasmonic mode. Thus, localized light fields were observed at near top surface of graphene layer (See \textcolor{blue}{Fig. S3} in \textcolor{blue}{Supplementary Information}). A single transparent window appeared at the valley of the Lorentz curve between two bright plasmonic modes. When x-polarized light was incident on entire structure, a transparent transmission window with transmission intensity $<$ 96\% emerge at the valley of Lorentz curve as can be seen in Fig.~\ref{fig3}(c). Transition states of the switch using different chemical potentials are listed in Table~\ref{Tab:1}. Chemical potentials of 0.2 eV and 0.5 eV were employed at two patterned graphene layers to achieve different transition states.
\begin{table}[H]
    \centering
    \caption{Chemical potentials of graphene layers for four state THz switch at different states}
    \begin{tabular}{ >{\centering\arraybackslash} m{1.5cm} >{\centering\arraybackslash} m {2cm} >{\centering\arraybackslash} m {2cm} >{\centering\arraybackslash} m {2cm}}
        \hline
        State & $\mu_{c_1}$ (eV)& $\mu_{c_2}$ (eV) & Illustration\\
        \hline
        00 & 0.5 & 0.2 & Fig.\,\ref{fig3}(c)\\
        \hline
        01 & 0 & 0.2 & Fig.\,\ref{fig3}(d)\\
        \hline
        10 & 0.5 & 0 & Fig.\,\ref{fig3}(e)\\
        \hline
        11 & 0  & 0 & Fig.\,\ref{fig3}(f)\\
        \hline
    \end{tabular}
    \label{Tab:1}
\end{table}
Figs.~\ref{fig3}(c)-(f) depict the various states of the THz switch. The chemical potentials were 0.5 eV and 0.2 eV, 0 eV and 0.2 eV, 0.5 eV and 0 eV, and 0 eV and 0 eV for OFF - OFF (00), OFF - ON (01), ON - OFF (10), and ON - ON (11) states, respectively. The OFF - OFF (00) state has two plasmonic bright modes at 3.357 THz and 5.4395 THz, while the ON - ON (11) state has two plasmonic dark modes at 3.357 THz and 5.4395 THz. To determine the performance of the THz switch, MD was calculated by~\cite{Zhimin},

\begin{equation}
    \mbox{MD}\ =\ \frac{T_{on}-T_{off}}{T_{on}}.
\end{equation}

Here, $T_{on}$ and $T_{off}$ denote the magnitude of transmittance in the on and off state, respectively. Our four state switch structure yielded MDs of 95.66\% and 98.81\% at resonant frequencies of 3.357 THz and 5.4395 THz, respectively.

\subsection{Eight State Switching Scheme}

\begin{figure}[H]
\centering\includegraphics[width=\textwidth]{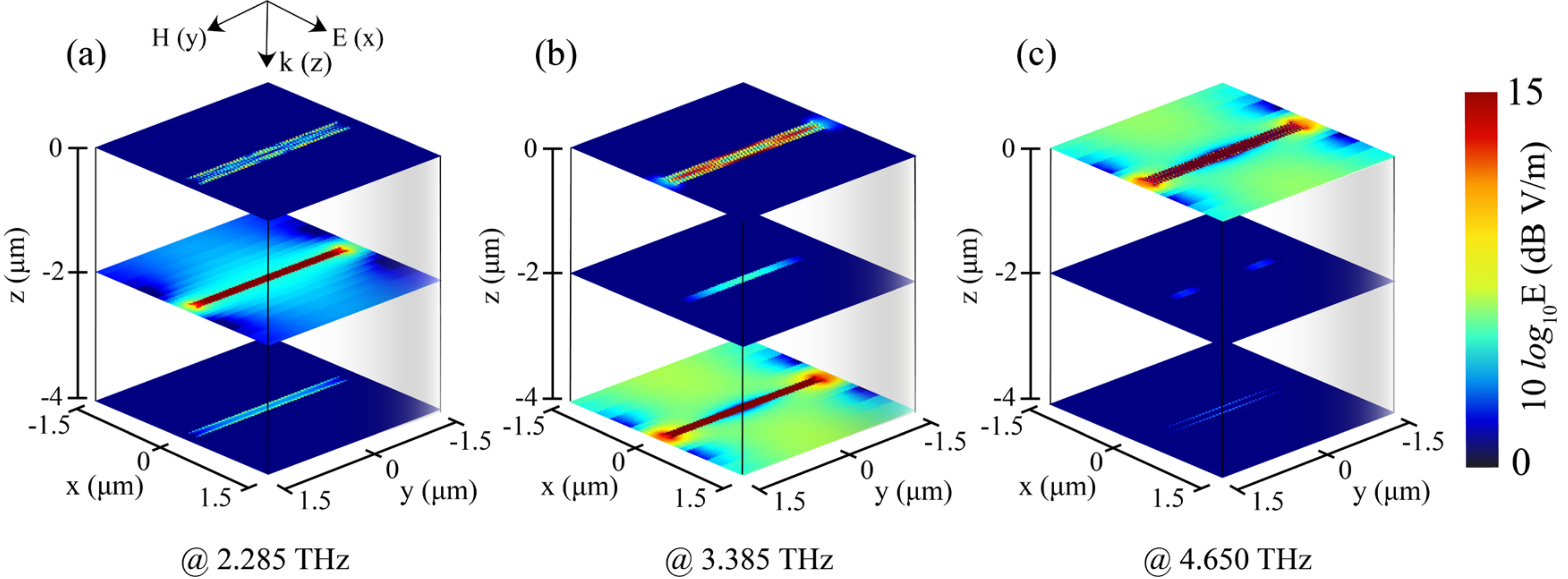}
\caption{The spatial E-field distribution of three patterned graphene layers at resonant frequencies of (a) 2.285 THz, (b) 3.385 THz, and (c) 4.65 THz for eight state THz switch. The E-field distributions are shown in log scale.}
\label{fig4}
\end{figure}

An eight state THz switch consisted of three patterned graphene layers on PMMA substrate was designed using localized plasmonic effect. Two transparent windows emerged at the valley of the Lorentz curves between three plasmonic bright modes due to surface conductivity of graphene layers using different chemical potentials. Transition states of eight state narrow band THz switch using different chemical potentials of graphene layers are listed in Table~\ref{Tab:3 state}. The chemical potentials of 0.2 eV, 0.3 eV, and 0.35 eV were employed via voltage sources by changing surface conductivity of graphene. Two transparent windows with maximum transmission intensity of $\sim$85\% at the valley of Lorentz curves appeared when an x-polarized light was incident from the top along the z direction. To analyze plasmonic effect extensively, the field distributions of three dips were delineated in Fig.~\ref{fig4}. Dominant localized electric fields were presented at the top, middle, and bottom graphene layers when the THz beam frequencies were 4.65 THz, 2.285 THz, and 3.385 THz, respectively (See \textcolor{blue}{Fig.\,S4} in \textcolor{blue}{Supplementary Information} for xz and yz cross-section of E-field distributions).

\begin{table}[H]
    \centering
    \caption{Chemical potentials of graphene layers for eight state THz switch at different states}
    \begin{tabular}{ >{\centering\arraybackslash} m{1.5cm} >{\centering\arraybackslash} m{2cm} >{\centering\arraybackslash} m{2cm} >{\centering\arraybackslash} m{2cm} >{\centering\arraybackslash} m{2cm}}
        \hline
        State & $\mu_{c_1}$ (eV) & $\mu_{c_2}$ (eV) & $\mu_{c_3}$ (eV) & Illustration\\
        \hline
        000 & 0.35 & 0.3 & 0.2 & Fig.\,\ref{fig5}(a)\\
        \hline
        001 & 0 & 0.3 & 0.2 & Fig.\,\ref{fig5}(b)\\
        \hline
        010 & 0.35 & 0.3 & 0 & Fig.\,\ref{fig5}(c)\\
        \hline
        011 & 0 & 0.3 & 0 & Fig.\,\ref{fig5}(d)\\
        \hline
        100 & 0.35 & 0 & 0.2 & Fig.\,\ref{fig5}(e)\\
        \hline
        101 & 0 & 0 & 0.2 & Fig.\,\ref{fig5}(f)\\
        \hline
        110 & 0.35 & 0 & 0 & Fig.\,\ref{fig5}(g)\\
        \hline
        111 & 0 & 0 & 0 & Fig.\,\ref{fig5}(h)\\
        \hline
    \end{tabular}
    \label{Tab:3 state}
\end{table}

\begin{figure}[H]
\centering\includegraphics[width=\textwidth]{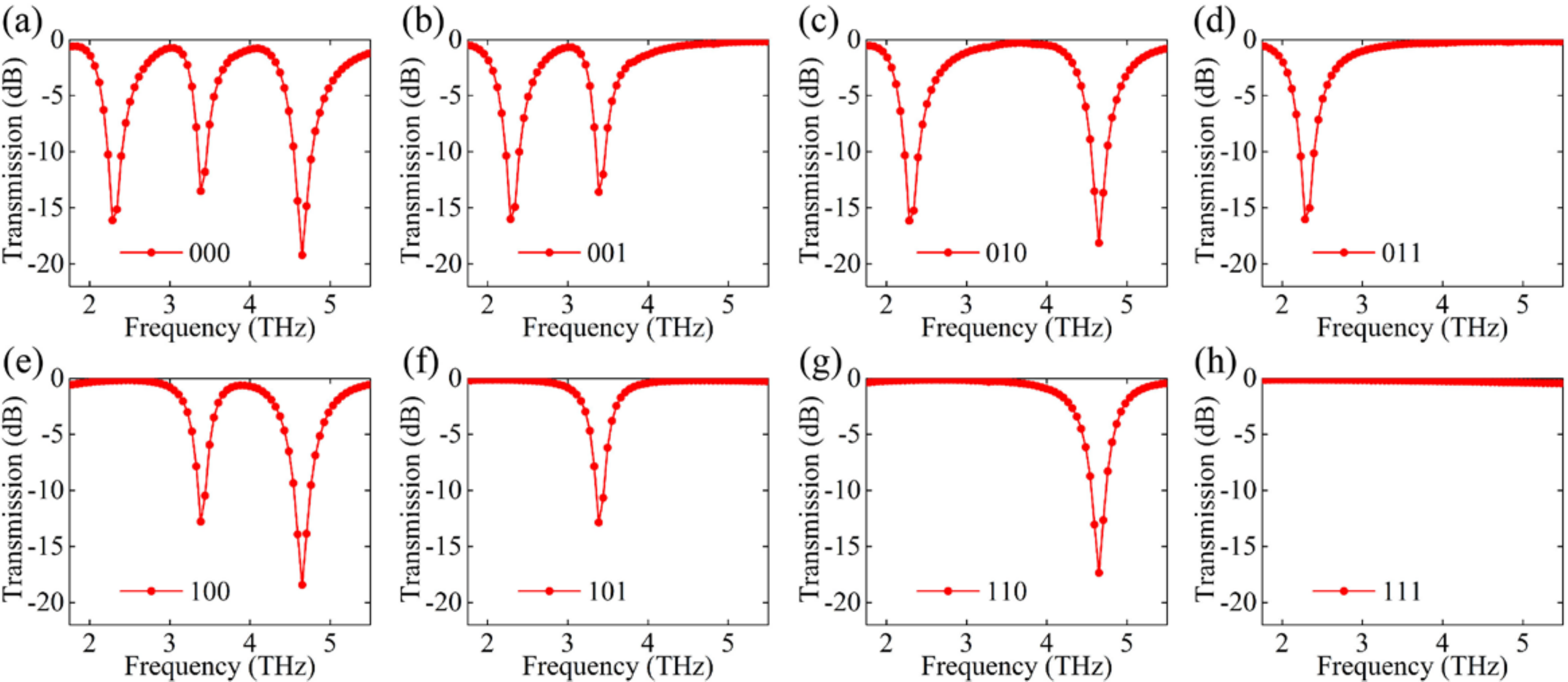}
\caption{Transmission spectra of eight state THz switch at different states: (a) “000” state, (b) “001” state, (c) “010” state, (d) “011” state, (e) “100” state, (f) “101” state, (g) “110” state, and (h) “111” state.}
\label{fig5}
\end{figure}

The transmission spectra of the eight state THz switch at all possible states are shown in Fig.~\ref{fig5}. Three plasmonic bright modes of the “000” state were appeared at resonant frequencies of 2.285 THz, 3.385 THz, and 4.65 THz. On the contrary, three plasmonic dark modes of the “111” state were existed at resonant frequencies of 2.285 THz, 3.385 THz, and 4.65 THz. The calculated MD was 97.47\%, 95.33\%, and 98.71\% at resonant frequencies of 2.285 THz, 3.385 THz, and 4.65 THz, respectively. Additionally, the cross sectional views of the field distributions for “010” and “111” states at three graphene patterned layers are provided in \textcolor{blue}{Supplementary Information (See Figs. S5-S8)}.
\subsection{Plasmon Tuning and Performance Analysis}
\begin{figure}[H]
\centering\includegraphics[width=\textwidth]{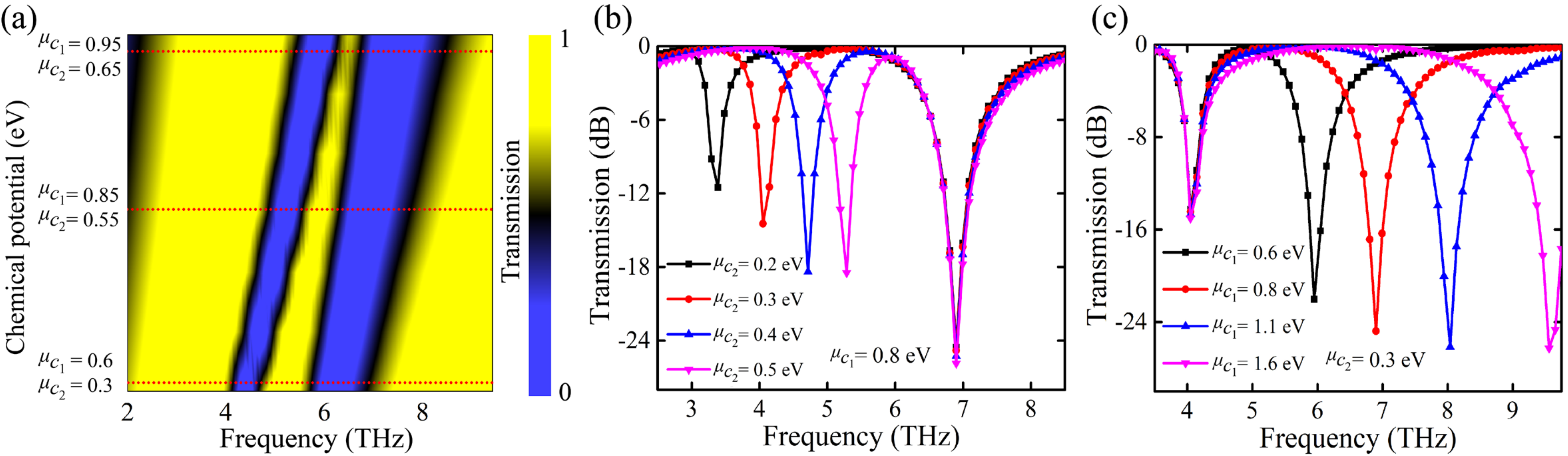}
\caption{(a) Evolution of plasmonic bright modes at different chemical potentials. (b)~Customization of the lower frequency dip (plasmonic bright mode) keeping $\mu_{c_1}$ at 0.8 eV. (c) Frequency tuning of the higher frequency dip (plasmonic bright mode) keeping $\mu_{c_2}$ at 0.3 eV.}
\label{fig6}
\end{figure}

\begin{figure}[H]
\centering\includegraphics[width=\textwidth]{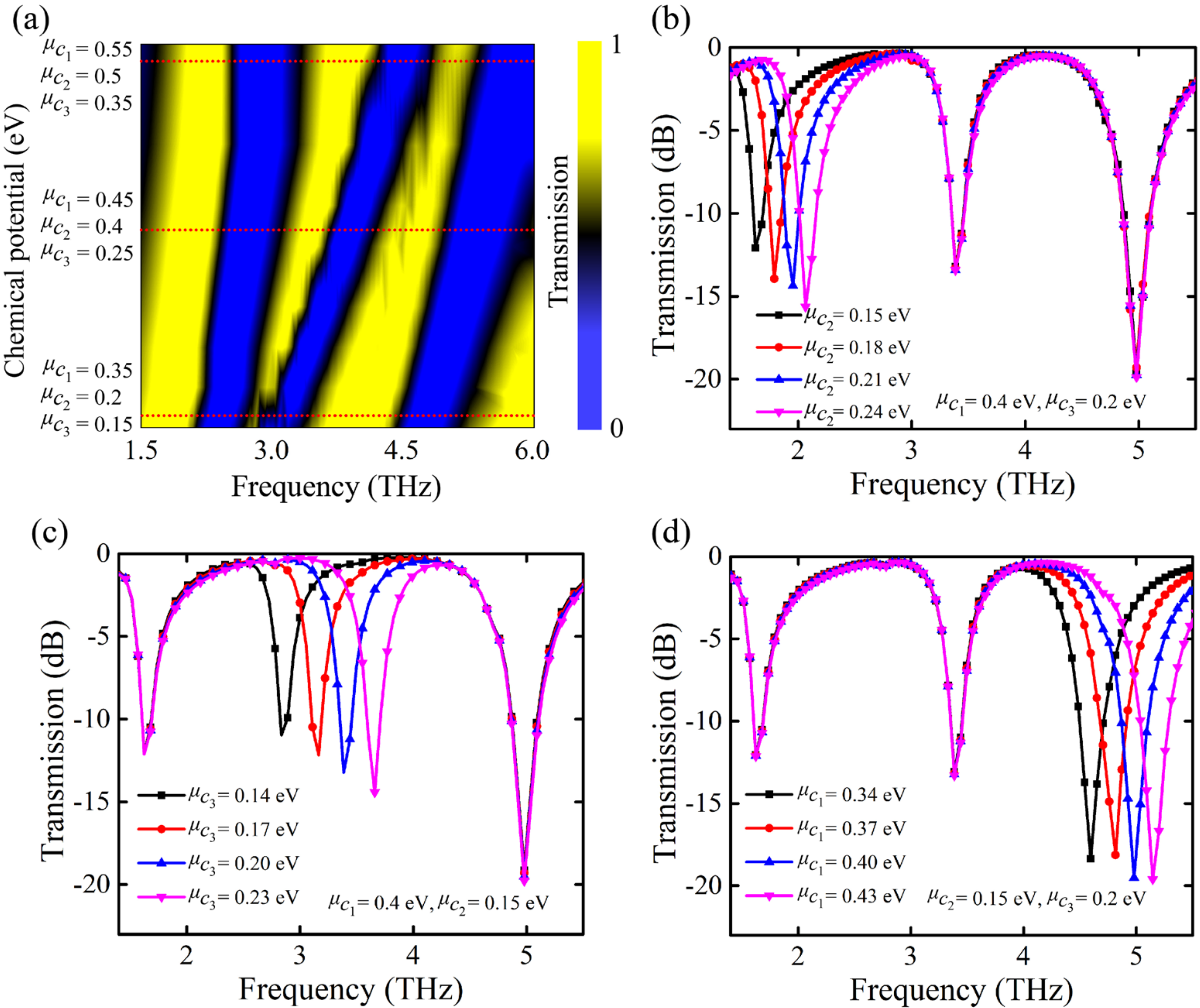}
\caption{(a) Evolution of plasmonic bright modes at different chemical potentials. (b)~Frequency variation of the first plasmonic bright mode considering $\mu_{c_1}$ = 0.4 eV and $\mu_{c_3}$ = 0.2 eV. (c) Frequency variation of the second plasmonic bright mode considering $\mu_{c_1}$ = 0.4 eV and $\mu_{c_2}$ = 0.15 eV. (d) Frequency variation of the third plasmonic bright mode considering $\mu_{c_2}$ = 0.15 eV and $\mu_{c_3}$ = 0.2 eV.}
\label{fig7}
\end{figure}

To tune the resonant frequency without any structural modification, the surface conductivity of graphene can be varied by changing $\mu_c$ in the wide range of 0 eV - 1.3 eV via appropriate external gate voltage V$_{\mbox{g}}$. Similar level of voltages were reported in previous studies\,\cite{Fallah2019,Zhang2020OE,Zhimin,Liu2020,Zhang2021}. The relationship of V$_{\mbox{g}}$ with $\mu_c$ is given by~\cite{ju2011graphene},
\begin{equation}
    \mu_c\approx\hbar V_F \sqrt{\frac{\pi \epsilon_{PMMA} {\mbox{V}}_{\mbox{g}}}{e\tau}}.
    \label{eq:8}
\end{equation}
Here, $\epsilon_{PMMA}$ is the permittivity of the PMMA substrate. By changing the $\mu_c$ of graphene layers using external voltage sources as shown in Figs.~\ref{fig:1}(a) and (b), the resonant frequency can be widely tuned in the THz frequency regime. Here, we expect that the parasitic impedance due to metal contact to be very small and hence, it was ignored. A color map of transmittance, shown in Fig.~\ref{fig6}(a), for four state THz switch at different chemical potentials displays the evolution of plasmonic modes. It is evident from the Fig.~\ref{fig6}(a) that the plasmonic bright modes in the transmission spectra are blue shifted with the increase of chemical potentials. The chemical potentials of the top and bottom graphene layers were varied separately as can be seen in Figs.~\ref{fig6}(b) and (c). Noteworthy, the lower frequency dip exhibited a blue shift when the chemical potential of top graphene layer was varied from 0.6 to 0.95 eV. Similarly, the higher frequency  dip showed blue shift when the chemical potential of bottom graphene layer was changed from 0.3 to 0.65 eV. The modulation degree of frequency, MDF a performance parameter that describes the frequency tailoring capability of the switch, can be expressed mathematically by~\cite{Liu2020},
\begin{equation}
    \mbox{MDF}\ =\ \frac{|f_{max}-f_{min}|}{f_{min}}.
\end{equation}
Here, $f_{max}$ and $f_{min}$ are the frequencies of maximum and minimum modulated dips of transmission spectra, respectively. The MDFs of first and second resonant dips were calculated to be $\sim$56\% and $\sim$61\%. Additionally, to evaluate contrast of transmission between the transparent region and the bright plasmon mode, we calculated spectral contrast ratio, $S_{con}$ given by~\cite{Yan2017}, 
\begin{equation}
    S_{con} =\frac{T_{transparent}-T_{bright}}{T_{transparent}+T_{bright}}.
\end{equation}
Where, $T_{transparent}$ and $T_{bright}$ are the average transmittance of transparent region(s) and plasmonic bright modes. $S_{con}$ of our proposed four state THz switch was 96.2\% which indicated impressive sensitivity of our proposed switch.
The dependencies of transmission at the transparent regions and plasmonic bright modes with respect to frequency are given in \textcolor{blue}{Supplementary Information} (see \textcolor{blue}{Fig. S9}). A visualization of evolution of plasmonic bright modes for eight state THz switch with the variation of chemical potentials are depicted in Fig.~\ref{fig7}(a). The plasmonic bright modes were blue shifted as can be seen in Figs.~\ref{fig7} (b)-(d). The calculated MDFs were 27.08\%, 29.1\%, and 11.97\% for the three bright modes. Calculated $S_{con}$ was 96.3\% for eight state THz switch. 
Insertion loss,  a important performance parameter of infrared optical switches, was calculated by $-10log_{10}(T_{on})$~\cite{zubair2016carbon}. Insertion loss was enumerated to be 0.22 dB and 0.33 dB for four and eight state THz switches, respectively. These values are substantially low compare to that of previously reported optical switches~\cite{chu2013active,li2020ultracompact,li2018graphene,granpayeh2018tunable}.  Moreover, we obtained extinction ratio, ER = $10log_{10}(T_{on}/T_{off})$ of 19.67 dB and 18.89 dB for four and eight state THz switches, respectively.
\begin{figure}[H]
\centering\includegraphics[width=\textwidth]{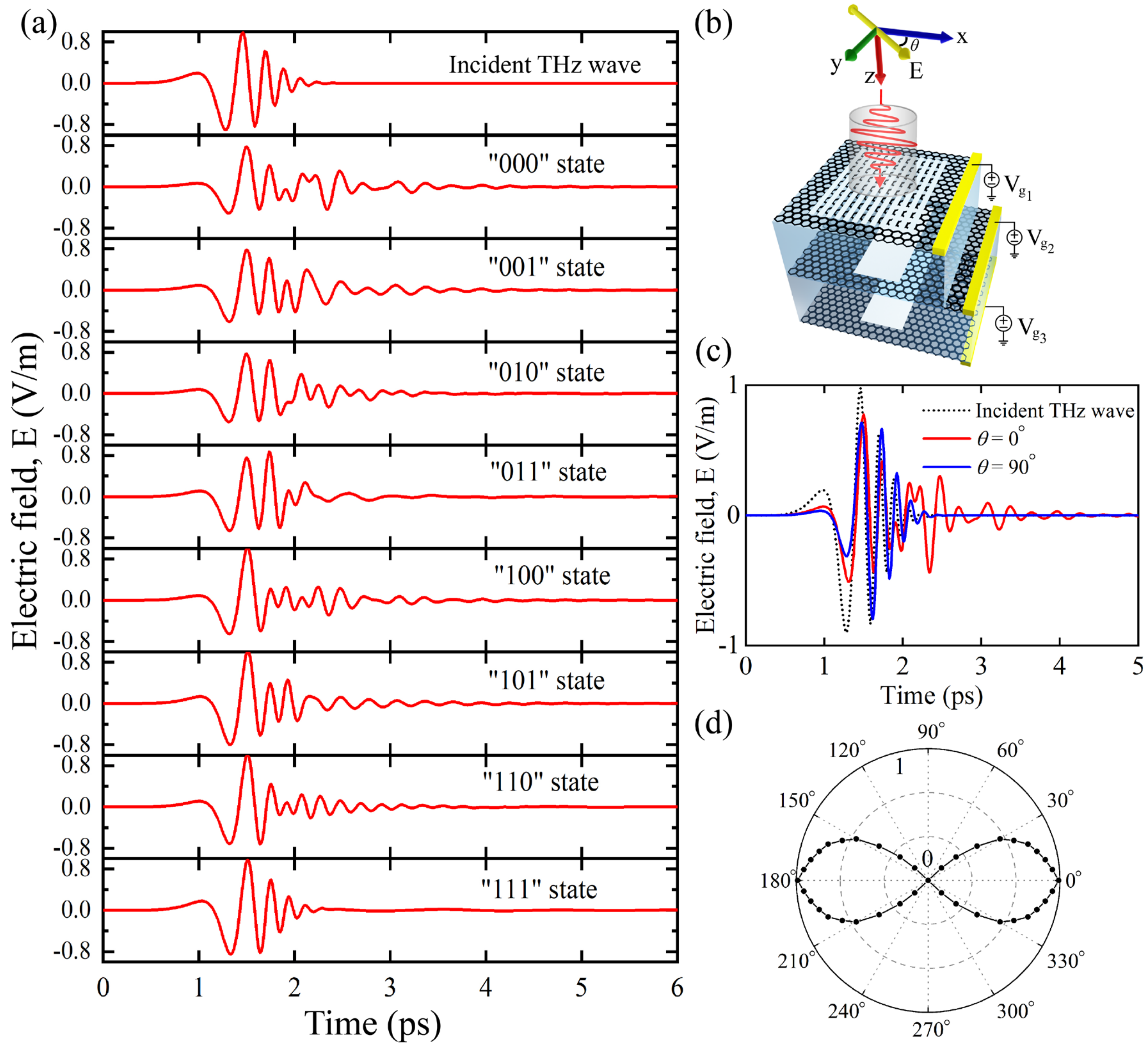}
\caption{(a) Time domain waveforms of incident and transmitted THz electric field for eight state optical switch at different states. (b) Illustration of polarized light incident on the eight state THz switch. (c) Time domain waveforms of transmitted THz electric field for eight state optical switch at $\theta$ = 0$^\circ$ and 90$^\circ$. (d) Polarization dependent normalized transmission of the eight state THz switch. The polarization angle, $\theta$ is an angle between x axis and the direction of the light polarization. In this polar plot, $\theta$ was swept from x polarized light to y polarized light.}
\label{fig8}
\end{figure}
 Furthermore, we  quantified the dephasing time, $\tau_{\phi}^{-1}$ ($= 2 \hbar/FWHM$)~\cite{Ahmadivand2016} for the induced deepest dip of our proposed devices. The estimated dephasing times were 2.27 ps and 1.87 ps for four and eight state THz switches, respectively. Fig.~\ref{fig8}(a) depicts transmitted THz wave in time domain for incident light and  different states of eight state THz switch. The electric field of “000” state attenuated the light at resonant frequencies. It is evident from Fig.~\ref{fig8}(a) that when the states were changed from “000” state to “111” state, the deviation of transmitted THz electric field reduced.
 
The polarization of the incident THz wave was varied for the eight state optical switch which is illustrated in Fig.~\ref{fig8}(b). It is evident from our calculation that all three plasmonic bright modes are polarization dependent (see \textcolor{blue}{Fig.~S10} in \textcolor{blue}{Supplementary Information}). The polarization angle is denoted by $\theta$ which is the angle between incident THz wave and x-axis. The time traces of incident wave for $\theta$ = 0$^\circ$ and $\theta$ = 90$^\circ$ are shown in Fig.~\ref{fig8}(c). Fig.~\ref{fig8}(d) delineates polar plot of normalized transmission ($(T_{ON}-T_{OFF})/T_{ON}$) at different polarization angles for eight state THz switch. We observed polarization dependency transmission with highest transmission for $\theta$ = 0$^\circ$ and lowest transmission for $\theta$ = 90$^\circ$. Hence, we can conclude that the presented THz optical switches are highly polarization sensitive.

\subsection{Comparative Analysis}
Table~\ref{Tab:3} demonstrates the comparison of performance parameters between our proposed THz switches and previously reported THz switches. A graphene ribbon array switch that operated in the mid-infrared regime, had the MD of 70\% along with high insertion loss~\cite{chu2013active}. Wei $et ~al.$ introduced a graphene THz metamaterial structure with considerably higher MD (96.2\%)~\cite{wei2016active}. This structure had narrower bandwidth than the array structure~\cite{chu2013active}. There are previous reports on  graphene micro-cavity~\cite{li2020ultracompact} and graphene-based non-volatile ~\cite{li2018graphene} THz switch that had similar bandwidth compared to our switches. However, MD and insertion loss were much lower.
\begin{table}[H]
    \centering
    \caption{Performance comparison of the proposed THz switches with previously reported works}
    \begin{tabular}{ >{\centering\arraybackslash} m{2.4cm} >{\centering\arraybackslash} m{1.6cm} >{\centering\arraybackslash} m{1.3cm} >{\centering\arraybackslash} m{1.1cm} >{\centering\arraybackslash} m{0.9cm}
    >{\centering\arraybackslash} m{0.9cm} >{\centering\arraybackslash} m{1cm}}
        \hline
        Description of structure & Operating wavelength regime & Bandwidth (THz) & MD (\%) & MDF (\%) & Insertion loss (dB) & Refe\-rence\\
        \hline
        Graphene ribbon array & Mid infrared & 1.87 & $<$ 70 & -- & 4.77 & ~\cite{chu2013active} \\
        
        Graphene metamaterial & Terahertz & 1.1 & $<$ 96.2 & -- & -- & ~\cite{wei2016active}\\
        
        Graphene micro-cavity & Terahertz & 0.2 & $<$ 90 & 30 & 0.36 & ~\cite{li2020ultracompact}\\
        
        Graphene based non-volatile & Terahertz & 0.12 & 89 & -- & 0.46 & ~\cite{li2018graphene} \\
        
        Grating structure & Infrared & -- & 92 & -- & 1.55 & ~\cite{granpayeh2018tunable} \\
        
        Four state switch & Terahertz & 0.14 & $<$ 98.81 & $\sim$ 61 & 0.22 & This work \\
        
        Eight state switch & Terahertz & 0.17 & $<$ 98.71 & 29.1 & 0.33 & This work \\
        \hline
        \label{Tab:3}
    \end{tabular}
\end{table}
The grating structure proposed by Khazaee $et ~al.$ had a decent MD of 92\%, but high insertion loss that operated in the infrared regime~\cite{granpayeh2018tunable}. In terms of MDF the graphene micro-cavity switch~\cite{li2020ultracompact} had 30\%, which is comparable to our eight state switch. But our four state switch had an MDF of almost double ($\sim$61\%). In comparison to all of the previously reported work, our proposed four and eight state switches had narrow bandwidth functioning in the THz regime with significantly higher MDs of 98.81\% and 98.71\% for both four and eight state switches, respectively.

\section{Conclusions}
In this work, we proposed and numerically analyzed plasmonic effect based four and eight state narrow band THz switches that had high MD, extremely low insertion loss, wide range MDF, and high spectral contrast ratio. Because of the strong interaction between patterned graphene layers and incoming light, plasmonic bright modes were originated. We exploited these interaction of plasmonic bright modes and dark mode in patterned multilayer of graphene on PMMA substrate to design the THz switches. The modulation of surface conductivity by varying chemical potentials of different graphene layers resulted in modulation of plasmonic bright modes. Consequently, narrow transparent band between bright modes emerged. High values of transmission intensities between two bright modes were obtained with maximum MDs of 98.81\% and 98.71\% for our proposed four and eight state THz switches, respectively. Interestingly, feasible wide tuning of the bright modes were demonstrated by adjusting the applied voltage of graphene metamaterial without any structural modification. We obtained exceedingly low insertion losses of 0.22 dB and 0.33 dB with wide MDFs. Moreover, the bandwidth at operating frequencies were as narrow as few hundreds of GHz. This study will put forward a basis for designing and manufacturing multimode plasmonic switches and modulators in THz frequency regime for communication and sensing applications.

\label{sec:conc}
\section*{Acknowledgements}
D. Sarker, P. P. Nakti, M. I. Tahmid, and M. A. Z. Mamun acknowledge the department of Electrical and Electronic Engineering (EEE), Shahjalal University of Science and Technology (SUST) for providing essential facilities for the completion of the study. D. Sarker, M.I. Tahmid, and A. Zubair  greatly appreciate the Department of EEE at Bangladesh University of Engineering and Technology (BUET) for technical assistance.
\\
\section*{Declarations}
\textbf{Ethical approval}\ \ \ Not applicable.\\
\\
\textbf{Competing interests}\ \ \ The authors declare no competing interests of a financial or personal nature.\\
\\
\textbf{Authors' contributions}\ \ \  D.S. and P.P.N. conducted the investigation, performed the analysis, wrote the main manuscript text, and prepared figures. D.S. and P.P.N. contributed equally to this work. M.I.T. and M.A.Z.M. conducted the investigation and wrote the main manuscript text. A.Z. supervised and administered the work and wrote the main manuscript text. All authors reviewed the manuscript.\\
\\
\textbf{Funding}\ \ \ Not applicable.\\
\\
\textbf{Availability of data and materials}\ \ \ Data and materials underlying the results depicted in this work are not publicly available at this moment but maybe obtained from the authors upon reasonable request.
\bibliography{bibtex}
\bibliographystyle{unsrt}


\end{document}